\documentclass{article}

\usepackage{graphicx}
\usepackage{latexsym}
\usepackage[fleqn]{amsmath}
\usepackage{amssymb}
\usepackage{amsbsy}
\usepackage{subcaption}
\usepackage{nishimura_arxiv}

\begin{document}

\title{Stochastic Safety-critical Control Compensating Safety Probability for Marine Vessel Tracking}

\author{Too Matsuo\footnotemark[1], Y\^uki Nishimura\footnotemark[2], Kenta Hoshino\footnotemark[3] and Daisuke Tabuchi\footnotemark[1]}
\date{March 30, 2026\footnotemark[3]}

\renewcommand{\thefootnote}{\fnsymbol{footnote}}
\footnotetext[1]{Kagoshima University}
\footnotetext[2]{Okayama University}
\footnotetext[3]{DENSO IT Laboratory}
\footnotetext[4]{This work has been submitted to the Journal of Marine Science and Technology for possible publication. Copyright may be transferred without notice, after which this version may no longer be accessible.}

\renewcommand{\thefootnote}{\arabic{footnote}}
\maketitle

\begin{abstract}
A marine vessel is a nonlinear system subject to irregular disturbances such as wind and waves, which cause tracking errors between the nominal and actual trajectories. In this study, a nonlinear vessel maneuvering model that includes a tracking controller is formulated and then controlled using a linear approximation around the nominal trajectory. The resulting stochastic linearized system is analyzed using a stochastic zeroing control barrier function (ZCBF). A stochastic safety compensator is designed to ensure probabilistic safety, and its effectiveness is verified through numerical simulations.
\end{abstract}

\section{Introduction}\label{sec:introduction}

Autonomous marine vessel control is an essential control application problem aimed at reducing labor shortages and maritime accidents.
Since vessel dynamics constitutes nonlinear control systems subject to disturbances such as wind and waves, their control problems are generally challenging, and various methods have been developed. The basic modelings of the system and control design procedures for the vessels are summarized in Fossen \cite{fossen}. 

Recently, control designs that stabilize vessels to follow a nominal trajectory are proposed based on nonlinear control theory \cite{fujii,saback}. Fujii et al. \cite{fujii} proposed a control design that stabilizes the vessel to track the nominal trajectory based on input-to-state stability, and Saback et al. \cite{saback} developed a control approach based on nonlinear model predictive control. These strategies achieve compensation for obstacle avoidance, modeling errors, and disturbance suppression, which are essential from practical perspectives. 

For more flexible achievement of safety and ease of control design, control designs based on safety-critical control theory are effective; see, for example Ames et al. \cite{ames2019}. Otsuki et al. \cite{otsuki} applies the safety-critical control theory into the tracking problem of a vessel by combining model predictive control with safety-critical control for guidance in port environments. 

At the same time, because marine vessels are systems subject to various irregular disturbances, safety under disturbances is required to compensate as in \cite{esfahani2021}. When irregular disturbances are regarded as stochastic fluctuations, the systems are sometimes considered as stochastic systems \cite{maki2023,maki2024}. Maki et al. \cite{maki2023} analyzes the stabilizability and destabilizability effects of the multiplicative stochastic fluctuations for ship's maneuvering motion. Maki et al. \cite{maki2024} analyzes the mechanism
of parametric rolling in irregular seas based on stochastic Lyapunov stability theory. However, to the best of the author's knowledge, the application of the recent results of stochastic safety-critical control theory to ship automation remains a challenging task.

Safety-critical control theory for stochastic systems are developed by Prajana et al. \cite{prajana}, Xue et al. \cite{xue} and Nejati et al. \cite{nejati} based on the stochastic safety verification. These methods are effective for complex safety-critical control problems, while the analyzing and control methods are complicated. An advantage of safety-critical control is its simple control design framework based on a control barrier function. Taking advantage of this feature, a stochastic safety-critical control theory is proposed by Nishimura and Hoshino \cite{nishimura:24}. Furthermore, an analytical method for safety probability compensation in stable linear systems is developed by Nishimura and Hoshino \cite{nishimura}. In the methods, the diffusion coefficient, which directly characterizes the influence of disturbances on the system, explicitly appears in the sufficient condition for safety. Therefore, in a control design that utilizes the sufficient condition, the diffusion coefficient directly influences the control law. At the same time, for the purpose of compensating for the influence of stochastic noises of tracking problems, the results in \cite{nishimura:24} and \cite{nishimura} should be improved a little.

In this paper, we slightly modify the stochastic safety-critical control theory in \cite{nishimura:24} and \cite{nishimura}, followed by formulating a vessel maneuvering model as a stochastic linear system subject to disturbances. Then, we apply the modified stochastic safety-critical control technique to a tracking problem of a vessel under irregular disturbances so that the vessel remains within a designed region with a specified probability. The validity of the designed compensator is confirmed by numerical simulation.

This paper is organized as follows. In Section~\ref{sec:model} , we describe our target system based on Fossen \cite{fossen}. In Section~\ref{sec:tracking}, we design a tracking controller using the linear quadratic (LQ) control scheme for the linearized system model. In Section~\ref{sec:analysis}, we analyze the safety probability, which is the probability that the error trajectory remains the designed region when Gaussian white noise is introduced into the linearized system. Then, in Section~\ref{sec:linear}, we design a linear compensator to increase the safety probability for the linearized system, and in Section~\ref{sec:nonlinear}, we also design a nonlinear compensator of the safety probability for the target nonlinear system. In Section~\ref{sec:simulation}, we confirm the validity of the designed compensators by numerical simulations. The safety-probability control theory is proposed in Section~\ref{sec:theory} by modifying the results of \cite{nishimura:24} and \cite{nishimura}, and in Section~\ref{sec:conclusion}, we conclude the paper.

\section{System Model}\label{sec:model}

 \begin{figure}[t]
 \centering
 \includegraphics[width=60mm]{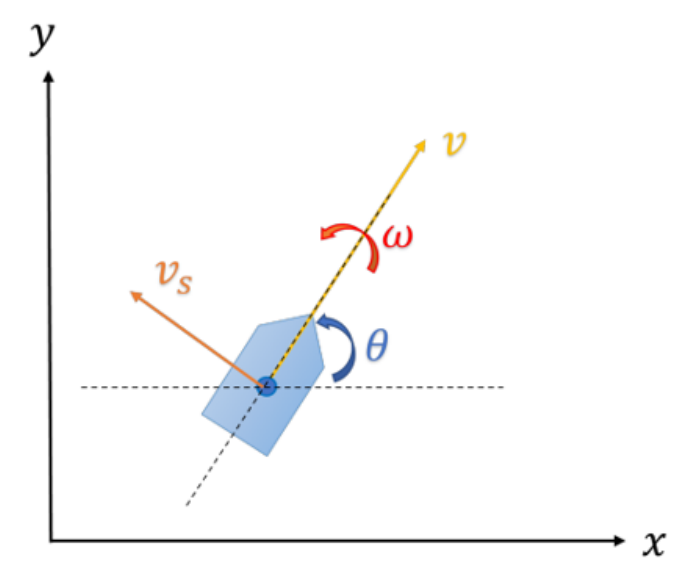}
 \caption{marine vessel maneuvering kinematic model \cite{fossen}.}
 \label{fig:vessel}
 \includegraphics[width=60mm]{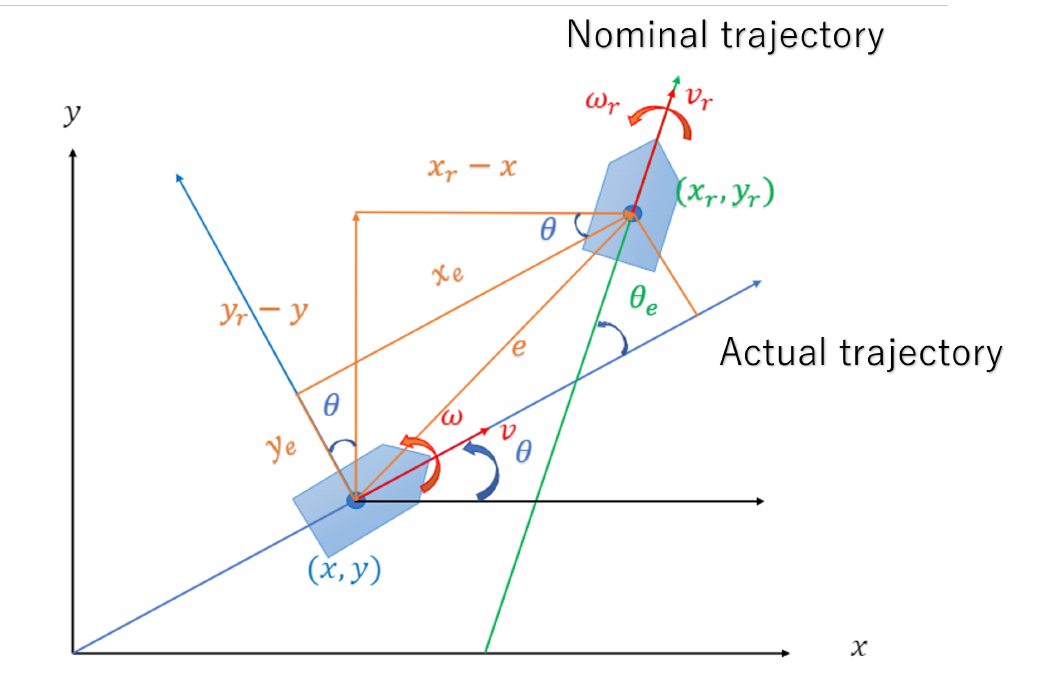}
 \caption{Marine vessel maneuvering kinematic model for nominal and actual trajectories.}
 \label{fig:vesseltrajectories}
 \end{figure}

In this section, we consider a marine vessel maneuvering kinematic model based on global position coordinates based on \cite{fossen}. 
Let the variables be defined as shown in Fig. \ref{fig:vessel}. Then we obtain 
\begin{align}\label{model_bef}
    \left\{ \begin{array}{lll}
         \dot{x}_{glo} = v_{sur} \cos\theta_{glo}-v_{swa}\sin\theta_{glo}, \\
         \dot{y}_{glo} = v_{sur} \sin\theta_{glo}+v_{swa}\cos\theta_{glo}, \\
         \dot{\theta}_{glo} = \omega_{yaw}, 
    \end{array}\right.
\end{align}
where $x_{glo}$ [m] is the horizontal position of the marine vessel, $y_{glo}$ [m] is vertical position of the marine vessel, $\theta_{glo}$ [rad] is the angle from the $x_{glo}$-axis, $v_{sur}$ [m/s] is surge velocity of the marine vessel, $v_{swa}$ [m/s] is sway velocity, and $\omega_{yaw}$ [rad/s] is yaw angular velocity.

In the same way as in \cite{fossen}, we modify the above model by pivoting on the point, which serves as the turning axis of the vessel.
If the pivoting point is located at a distance $c$ [m] ahead of the vessel’s center of gravity, we have
\begin{align}
\label{pivot}
c = -\frac{v_{swa}}{\omega_{yaw}},
\end{align}
where we assume $c\neq -1$. By substituting \eqref{pivot} to \eqref{model_bef}, we obtain a state-space model
\begin{align}\label{model_aft}
    \left\{ \begin{array}{lll}
         \dot{x}_{glo} = v_{sur}\cos\theta_{glo} + c\omega_{yaw} \sin\theta_{glo}, \\
         \dot{y}_{glo} = v_{sur}\sin\theta_{glo} - c\omega_{yaw} \cos\theta_{glo}, \\
         \dot{\theta}_{glo} = \omega_{ywa}. 
    \end{array}\right.
\end{align}

Here, defining the reference states $x_r$, $y_r$, and $\theta_r$ and the reference control inputs $v_r$ and $\omega_r$ with $v_r\neq 0$,  the reference model for \eqref{model_aft} is described as follows: 
\begin{align}\label{modelr}
    \left\{ \begin{array}{lll}
         \dot{x}_r=v_r\cos\theta_r+c\omega_r \sin\theta_r, \\
         \dot{y}_r=v_r\sin\theta_r-c\omega_r \cos\theta_r, \\
         \dot{\theta}_r=\omega_r. 
    \end{array}\right.
\end{align}

We define the tracking errors with respect to the reference states by $x_e, y_e, \theta_e$.
From the geometric relations illustrated in Fig. \ref{fig:vesseltrajectories}, these errors are given by
\begin{align}\label{modele}
    \left\{ \begin{array}{lll}
         x_e=(x_r-x_{glo})\cos\theta_{glo} + (y_r-y_{glo})\sin\theta_{glo}, \\
         y_e=-(x_r-x_{glo})\sin\theta_{glo} + (y_r-y_{glo}) \cos\theta_{glo}, \\
         \theta_e=\theta_r-\theta_{glo}. 
    \end{array}\right.
\end{align}
Differentiating both terms of the above, the error dynamics is derived as
\begin{align}\label{modeledot_pre}
    \left\{ \begin{array}{lll}
         \dot{x}_e=v_r\cos\theta_e+c\omega_r \sin\theta_e-v_{sur}+y_e\omega_{yaw}, \\
         \dot{y}_e=v_r\sin\theta_e-c\omega_r \cos\theta_e+(c-x_e)\omega_{yaw}, \\
         \dot{\theta}_e=\omega_r-\omega_{yaw}. 
    \end{array}\right.
\end{align}
Moreover, we set new control inputs $v$ and $\omega$ and design
\begin{align}
    &v_{sur} = v_r \cos(\theta_e) + v, \\
    &\omega_{yaw} = \omega_r \cos(\theta_e) + \omega,
\end{align}
so that the right-hand sides of \eqref{modeledot_pre} become all zero as $x_e=y_e=\theta_e=0$. Then, we obtain
\begin{align}\label{modeledot}
   &\dot{x}(t)= f(x(t))+ g(x(t))u(t), 
\end{align}
where $x = (x_e,y_e,\theta_e)^T$ is a state vector, $u=(v,\omega)^T$ is a control input vector, and
\begin{align}
    &f(x)=\begin{bmatrix}
        c\omega_r\sin\theta_e + \omega_r y_e \cos \theta_e \\
        v_r\sin\theta_e - \omega_r x_e \cos\theta_e\\
        \omega_r (1 - \cos \theta_e)
    \end{bmatrix}, \\
    &g(x)=\begin{bmatrix}
        -1 & y_e \\
        0  & c-x_e \\
        0  & -1
    \end{bmatrix}.
\end{align}

Finally, assuming that Gaussian white noise (roughly, $dw(t)/dt$, where $w$ is a one-dimensional standard Wiener process) is imposed on the above model, we obtain the following stochastic system:
\begin{align}\label{dXt}
   &dx(t)= \{ f(x(t))+ g(x(t))u(t) \} dt + G dw(t), 
\end{align}
where
\begin{align}\label{det:G}
    G=
    \begin{bmatrix}      
        \sigma_x\\\sigma_y\\\sigma_{\theta}
    \end{bmatrix}
\end{align}
with $\sigma_x, \sigma_y, \sigma_\theta$ being constants. Moreover, we assume that $g(x)^T G \neq 0$ is satisfied regardless of the value of $x$.

\begin{remark}
The reason for employing the stochastic differential equation \eqref{dXt} is that the Wiener process $w(t)$ is theoretically not differentiable almost everywhere, while the Gaussian white noise is formally derived as $dw/dt$. 
\end{remark}

\section{Control Design for Trajectory Tracking}\label{sec:tracking}

In this section, we design a trajectory tracking control law under the assumption of no noise. Because the target system \eqref{modeledot} is a nonlinear system, assuming that the initial state $x(0)$ is not so large, we consider the following linearized model:
\begin{align}\label{linear}
    &\dot{x}(t) = A x(t) + B u(t),
\end{align}
where
\begin{align}
    &A= \pfrac{f}{x}\Big|_{x=0} = \begin{bmatrix} 0 & \omega_r & cw_r \\ -\omega_r & 0 & v_r \\ 0 & 0 & 0 \end{bmatrix},\ B= g(0) = \begin{bmatrix} -1 & 0  \\ 0 & c \\ 0 &-1
    \end{bmatrix}.
\end{align}
Since we assume $c\neq -1$ and $v_r\neq 0$, $(A,B)$ is controllable.

Here, we design the state-feedback law $u(t) = u_{tra}(x(t))$ so that the trajectory tracking is achieved. Considering a linear state-feedback law
\begin{align}\label{u_o}
    u_{tra}(x) = -Kx,
\end{align}
where $K$ is a $2 \times 3$ matrix, if there exist $3 \times 3$ positive definite symmetric matrices $P$ and $Q$ satisfying the Lyapunov equation
\begin{align}\label{APQ}
P\bar{A} + \bar{A}^T P = -Q
\end{align}
with $\bar{A} = A-BK$, then $x=0$ of the system \eqref{linear} with \eqref{u_o} is asymptotically stable.

To determine the feedback gain $K$, we design $u_{tra}(x)$ as a linear-quadratic (LQ) control law; that is,
\begin{align}\label{eq:lqr}
K = -R^{-1}B^TPx,
\end{align}
which minimizes the following cost
\begin{align}
    J(x,u) = \int_{0}^{\infty}(x^T Q' x + u^ T R u)dt.
\end{align}
The matrix $P$ is the positive definite solution of the following algebraic Riccati equation:
\begin{align}
    PA+A^TP-PBR^{-1}B^TP+Q'=0,
\end{align}
which results in \eqref{APQ} with \eqref{eq:lqr} and $Q = Q' + PBR^{-1}B^TP$. 

\section{Safety Probability Analysis}\label{sec:analysis}
In this section, we consider the situation of the existence of noise for the linearized model \eqref{linear}, and derive the safety probability that remains within a safe set against the noise. 

Consider the linearized system with a linear state-feedback law and Gaussian white noise; that is, \eqref{linear} with $u = u_{tra}(x)$ and adding the diffusion term with $G$ in \eqref{det:G}, then we obtain
\begin{align}\label{dx_Abar}
    dx(t) = \bar{A} x(t) dt + G dw(t).
\end{align}

When there is no noise, tracking is achieved through the LQ control. However, when noise is introduced into the system, the tracking error $x$ may become large. Therefore, we design the acceptable error range and analyze the probability that $x$ remains within this range. A simple way to determine the range is to consider a sublevel set of the Lyapunov function $V(x)=x^T P x$; that is, $\{x | V(x) \le M \}$ for some $M>0$, because it is an invariant set when there is no noise for any $M$. Therefore, we define
\begin{align}
    &h(x)=-x^T P x + M    \label{h}
\end{align}
and 
\begin{align}
    &\chi=\{x|h(x)>0\}\label{chi} = \{ x | V(x) < M \},
\end{align}
where the function $h(x)$ is said to be a stochastic zeroing control barrier function (stochastic ZCBF) and $\chi$ is said to be a safe set. The reason for defining $\chi$ as an open set and using $h(x)$ is to consider stochastic safety-critical control theory, which is proposed in Section~\ref{sec:theory} below. 

Then, consider Theorem~\ref{THM:LINEAR} (in Section~\ref{sec:theory}). Let $\mu \in (0,M)$ and 
\begin{align}\label{eq:b}
b = \frac{\mathrm{eigmin}[Q] - \mathrm{eigmin}[P]\frac{  \mathrm{tr}[G^T P G]}{M - \mu}}{2\mathrm {eigmax} [PGG^TP]},
\end{align}
where $\mathrm{tr}[X]$ is the trace, $\mathrm{eigmax}[X]$ is the maximum eigenvalue, and $\mathrm{eigmin}[X]$ is the minimum eigenvalue of a positive semi-definite matrix $X$, respectively. If $b>0$; that is,
\begin{align}\label{Mmu}
    M-\mu > \mathrm{tr}[G^T P G]\frac{\mathrm{eigmin}[P]}{\mathrm{eigmin}[Q]}
\end{align}
is satisfied for some $\mu \in (0,M)$, the system \eqref{dx_Abar} is safe in $(\chi_{h>\mu},\chi,1-e^{-b\mu})$, which means that, if the initial value $x(0)$ is in the initial state set $\chi_{h>\mu} = \{x | h(x) > \mu\} = \{x | V(x) \le M-\mu \}$, the state $x(t)$ remains the safe set $\chi$ for all $t \ge 0$ at least the probability of $1-e^{-b \mu}$. 

\begin{figure}[t]
 \centering
 \includegraphics[width=0.9\linewidth, keepaspectratio]{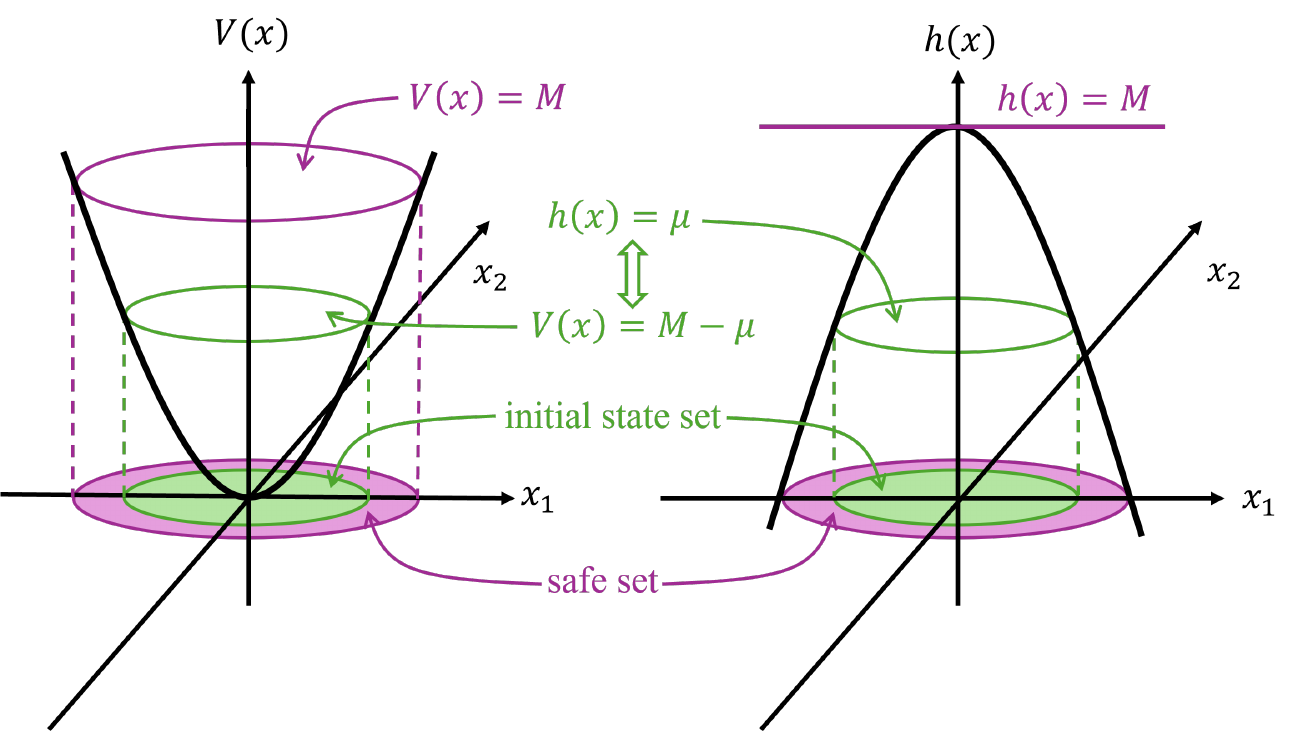}
 \caption{The brief sketch of the Lyapunov function $V(x)$, the stochastic ZCBF $h(x)$, the safe set $\chi$ and the initial state set $\chi_{h>\mu}$ for a state space $(x_1,x_2)^T$.}
 \label{fig:barrier}
\end{figure}

The specific image for the relationship between the Lyapunov function $V(x)$, the stochastic ZCBF $h(x)$, the safe set $\chi$ and the initial state set $\chi_{h>\mu}$ is drawn in Fig.~\ref{fig:barrier}, provided that the state space is reduced to two dimensions $(x_1,x_2)^T$ to simplify drawings.

\section{Stochastic Safety-critical Compensation by Linear State Feedback}\label{sec:linear}

In this section, we design a linear compensator $u_{com}(x(t)) = - K' x$ for the stabilized system \eqref{dx_Abar} to increase the probability that the error state vector $x$ keeps staying at $\chi$. To do this, we consider the system \eqref{linear} with $u= u_{tra}(x) + u_{com}(x)$ and \eqref{u_o} under the additive noise; that is,
\begin{align}\label{dx_Abar_com}
    dx(t) = (\bar{A} x(t) + B u_{com}(x(t)) ) dt+G dw(t). 
\end{align}
We also consider the stochastic ZCBF $h(x)$ in \eqref{h} and the safe set $\chi$ in \eqref{chi}.

Set
\begin{align}\label{eq:ucom}
    u_{com}(x) = - R'^{-1} B^T P x,
\end{align}
where $R'$ is a $2 \times 2$ positive definite and symmetric matrix. Then, according to Corollary~\ref{COR:SAFE-LIN-CLO2} (in Section~\ref{sec:theory}), we achieve that the system \eqref{linear} with $u= u_{tra}(x) + u_{com}(x) = -(R^{-1}+R'^{-1}) B^T P x$ is safe in $(\chi_{h>\mu},\chi,1-e^{-(b+b')\mu})$ with
\begin{align}\label{bplus}
    b^+ = \frac{G^T B R' B^T G}{G^TG}.
\end{align}
Because we assume $B^T G \neq 0$, $b^+ > 0$ is satisfied. Thus, the addition of $u_{com}(x)$ increases the safety probability; that is, $1-e^{-(b+b')\mu} > 1-e^{-b\mu}$.

\section{Stochastic Safety-critical Compensation by Nonlinear State Feedback}\label{sec:nonlinear}

In this section, we consider the nonlinear system \eqref{dXt} with $u(t) = u_{tra}(x) + u_{nlc}(x)$, where $u_{nlc}(x)$ is a nonlinear compensator. 

If the initial value of the state $x(0)$ is not so large, the behavior of the system \eqref{dXt} is approximated by the linearized system \eqref{linear}. If the noise is not very large, the linear compensator $u_{com}(x)$ achieves high safety probability as stated in the previous section. However, because we consider Gaussian white noise, the impact of noise could become extremely large instantaneously, and in such cases, the state value may also become large. Therefore, in this section, we design a nonlinear compensator for the nonlinear system \eqref{dXt}.

Letting a design parameter $b' >0$ and 
\begin{align}
    \gamma(x,b) = 2x^TP (f(x) +g(x) u_{tra}(x) + b' G G^T P x ) + G^T P G,
\end{align} 
we design $\phi_s(x)$ as follows: if $\gamma(x,b) >0$ and $x^T P g(x) \neq 0$,  
\begin{align}\label{phi_s}
\phi_s(x)
&= -u_{tra}(x) \notag\ \\
&\quad-\frac{2x^TP ( f(x) + b' G G^T P x) + G^T P G}
{2 x^T P g(x) g^T(x) P x} g^T(x) P x;
\end{align}
otherwise, $\phi_s(x)=0$.

Moreover, we consider
\begin{align}\label{phi}
u_{nlc}(x):=\left\{ \begin{array}{lll}
\phi_s(x),&h(x)\leq \mu\\
\phi_s'(x),\ \ & h(x)\in (\mu, M') \\
0, &h(x)\geq M'
\end{array}\right.
\end{align}
and 
\begin{align}\label{Phi_s}
    \phi_s'(x) = \phi_s(x)\frac{h(x)-M'}{\mu-M'}, 
\end{align}
where $M' \in (\mu,M]$ is designed so that $u_{nlc}=0$ for all $x \in \chi_{h > M'} \subset \chi_{h>\mu}$. 

Using Corollary~\ref{COR:CTRL-SZCBF}, we see that the system \eqref{dXt} with control input $u=u_{nlc}(x)$ is safe in $(\chi_{h>\mu},\chi,1-e^{-b'\mu})$. Moreover, if
\begin{align}
    2x^T P (f(x) + b' G G^T Px) + G^T P G > 0
\end{align}
holds for all $x\in\chi_\mu$ with some $\mu>0$ satisfying $x^T P g(x) = 0$, $u_{nlc}(x)$ is continuous regardless of the value of $x$. 

\section{Numerical Simulation}\label{sec:simulation}

\subsection{Parameter Settings}
In this section, we confirm the validity of the safety-probability compensators by calculating concrete problem settings and their numerical simulation. 

We first consider to add the LQ controller $u_{tra}(x)$ with \eqref{u_o} without any safety-probability compensator; that is, $u= u_{tra}(x)$. Setting $c=3$, $\omega_r=0.1$, and $v_r=1$, we obtain
\begin{align}
    A=\begin{bmatrix}
       0&0.1&0.3\\
       -0.1&0&1\\
       0&0&0
   \end{bmatrix},\quad
   B=  \begin{bmatrix}
        -1&0\\
        0&3\\
        0&-1
    \end{bmatrix}, 
\end{align}
and by setting
\begin{align}
    Q'=\begin{bmatrix}
        0.1&0&0\\
        0&0.3&0\\
        0&0&0.2
    \end{bmatrix},\quad
    R=\begin{bmatrix}
        40&0\\
        0&40
    \end{bmatrix},
\end{align}
the matrix $P$ is obtained as
\begin{align}
    P=\begin{bmatrix}
      1.99&-0.06&-0.92\\
     -0.06&2.64&11.32\\
    -0.92&11.32&63.81
    \end{bmatrix}.
\end{align}
Thus, the feedback gain $K$ is obtained as
\begin{align}
   K=\begin{bmatrix}
      -0.05&0.00&0.02\\
     0.02&-0.09&-0.75
    \end{bmatrix}
\end{align}
and the $\bar{A} = A - BK$ is given by
\begin{align}
  \bar{A}=\begin{bmatrix}
      -0.05&0.10&0.32\\
     -0.16&0.25&3.24\\
    0.02&-0.08&-0.75
    \end{bmatrix}.
\end{align}

Moreover, the diffusion coefficient $G$ is assumed to be 
\begin{align}\label{simG}
    G = \begin{bmatrix}
        0.08 \\ 0.08 \\ 0.08
    \end{bmatrix} .
\end{align}
To satisfy the condition \eqref{Mmu}, $M-\mu > 3.53$ is needed. Noting the condition, we set $M=10$ and $\mu=1$; then, we obtain $b = 0.0043$, and the system \eqref{linear} with $u=u_{tra}(x)$ is safe in $(\chi_{h>\mu}, \chi, 0.0043)$.


Next, to enhance the safety of the safe set, we consider to design the linear safety-probability compensator $u_{com}(x)$ given by \eqref{eq:ucom}. By choosing
\begin{align}
    R'=\begin{bmatrix}
        15&0\\
        0&15
    \end{bmatrix},
\end{align}
we obtain $b^+=5.79$ from \eqref{bplus}. Thus, the system \eqref{linear} with $u=u_{tra}(x)+u_{com}(x)$ is safe in $(\chi_{h>\mu},\chi,0.997)$.

Finally, we employ the nonlinear safety-probability compensator $u_{nlc}(x)$ given by \eqref{phi} and apply $u=u_{tra}(x) + u_{nlc}(x)$ to the system \eqref{dXt}. Setting $b'=3$ and $M'=9$, the system is safe in $(\chi_{h>\mu},\chi,0.950)$. 
 
\subsection{Simulation Results}

In this subsection, we compare the safety-probability performances of the three controllers $u=u_{tra}(x)$, $u=u_{tra}(x) + u_{com}(x)$ and $u=u_{tra}(x) + u_{nlc}(x)$ for the target nonlinear stochastic system \eqref{dXt} through numerical simulations.

For all simulations, we set the initial state vector $x(0)=(0.5,0.5,0)^T$, and thus $h(x_0)=8.87$. For each condition, we perform ten simulations with $G$ in \eqref{simG} and one simulation with $G=0$.

\begin{figure}[t]
 \centering
   \includegraphics[width=0.9\linewidth, keepaspectratio]{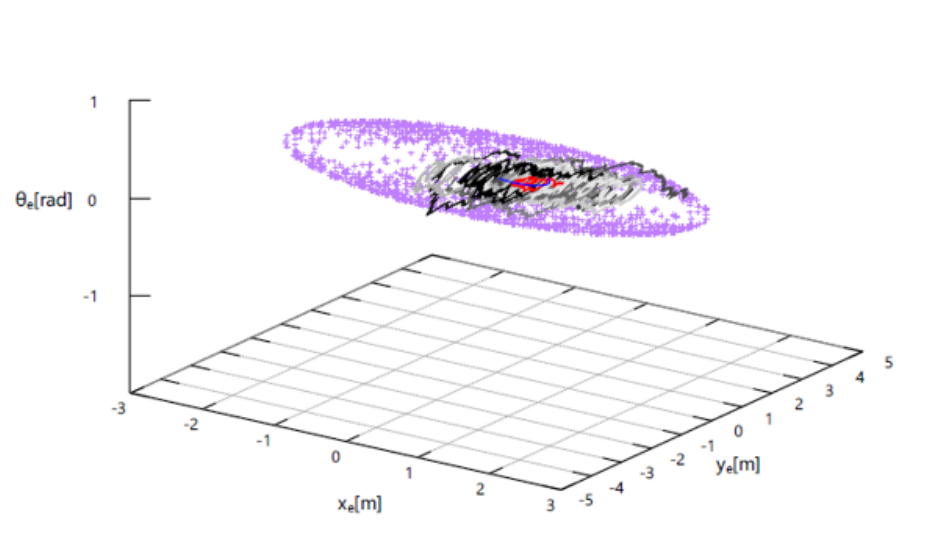}
   \caption{Trajectories with $u=u_{tra}$ and safety boundary.}
   \label{fig:noinp:f1a}
   \includegraphics[width=0.9\linewidth, keepaspectratio]{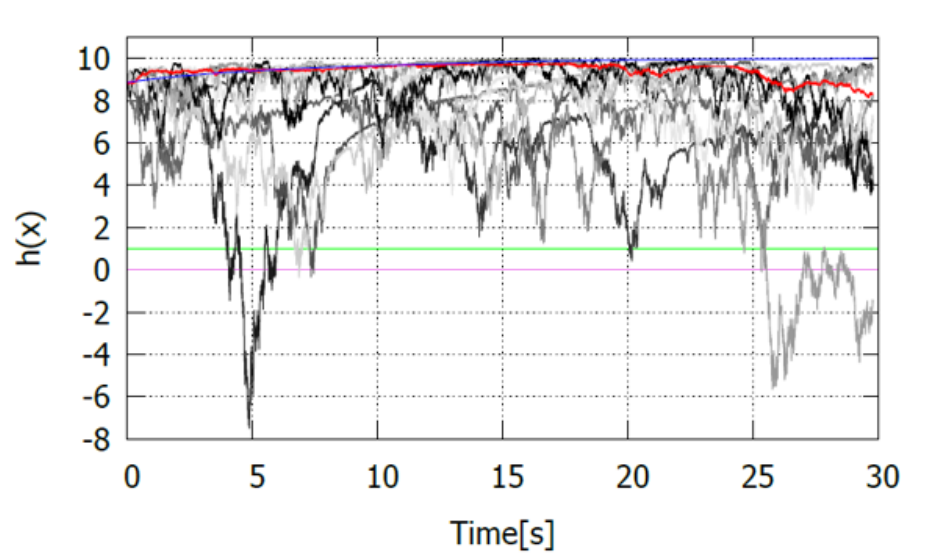}
   \caption{Time responses of $h(x)$ with $u=u_{tra}$.}
   \label{fig:noinp:f1b}
\end{figure}

\begin{figure}[t]
\centering
  \includegraphics[width=0.9\linewidth]{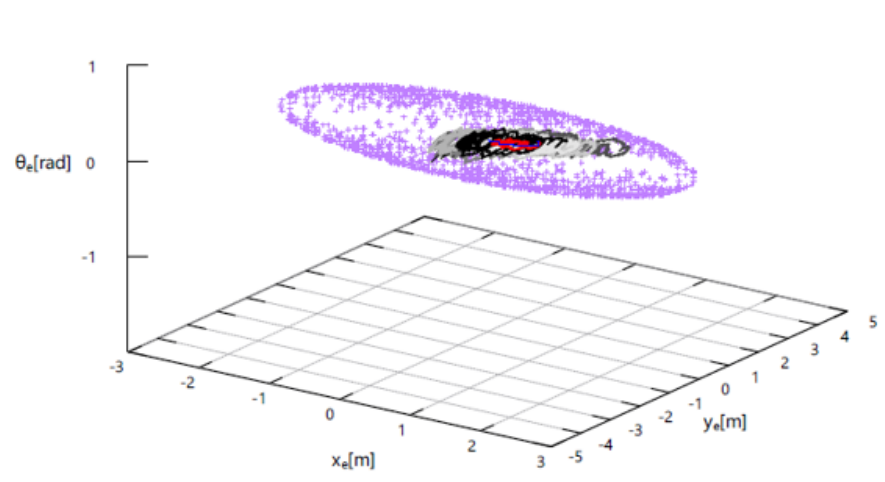}
  \caption{Trajectories with $u=u_{tra}+u_{com}$ and safety boundary.}
  \label{fig:withinp:f2a}
  \includegraphics[width=0.9\linewidth]{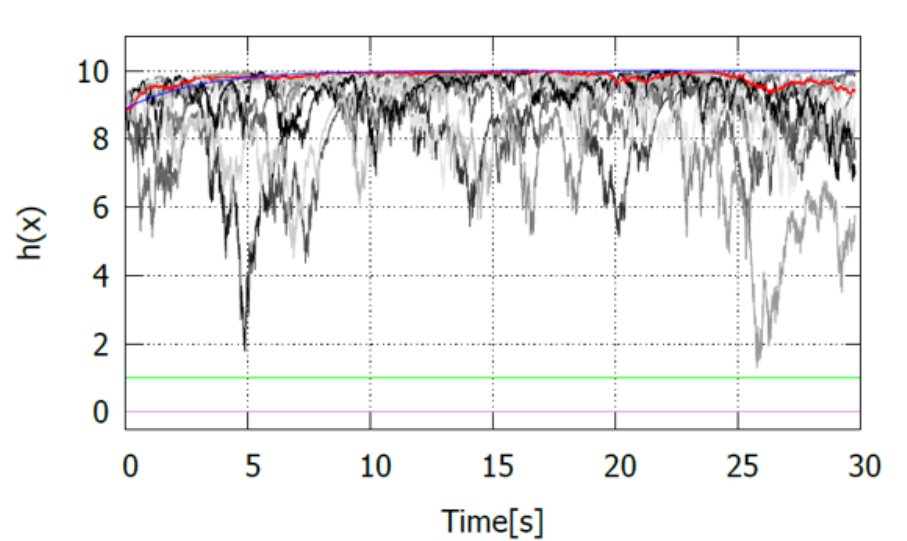}
  \caption{Time responses of $h(x)$ with $u=u_{tra}+u_{com}$.}
  \label{fig:withinp:f2b}
  \end{figure}

\begin{figure}[t]
  \centering
  \includegraphics[width=0.9\linewidth]{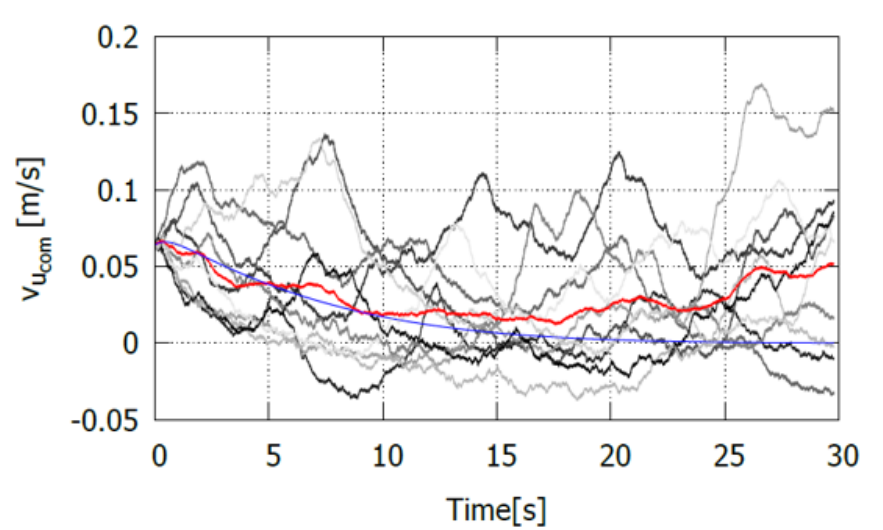}
  \caption{Time responses of $v_{com}$.}
  \label{fig:withinp:f2c}
  \includegraphics[width=0.9\linewidth]{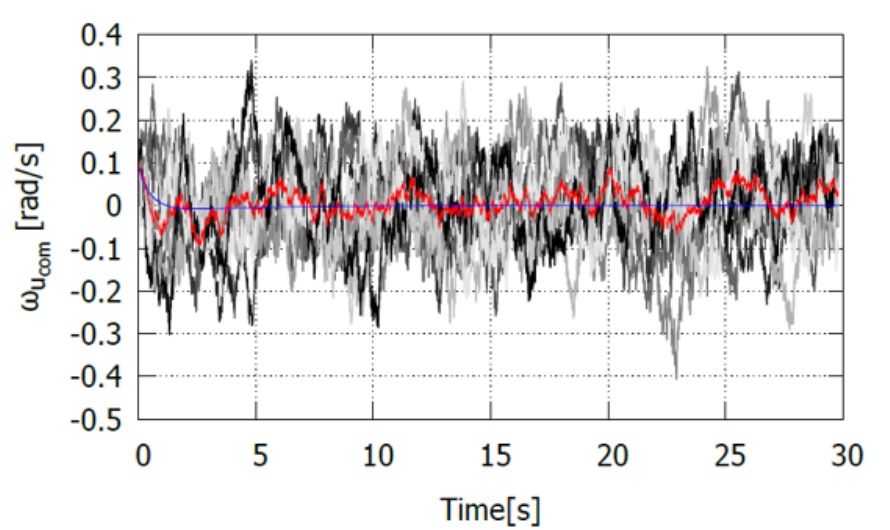}
  \caption{Time responses of $\omega_{com}$.}
  \label{fig:withinp:f2d}
\end{figure}

\begin{figure}[t]
\centering
  \includegraphics[width=0.9\linewidth]{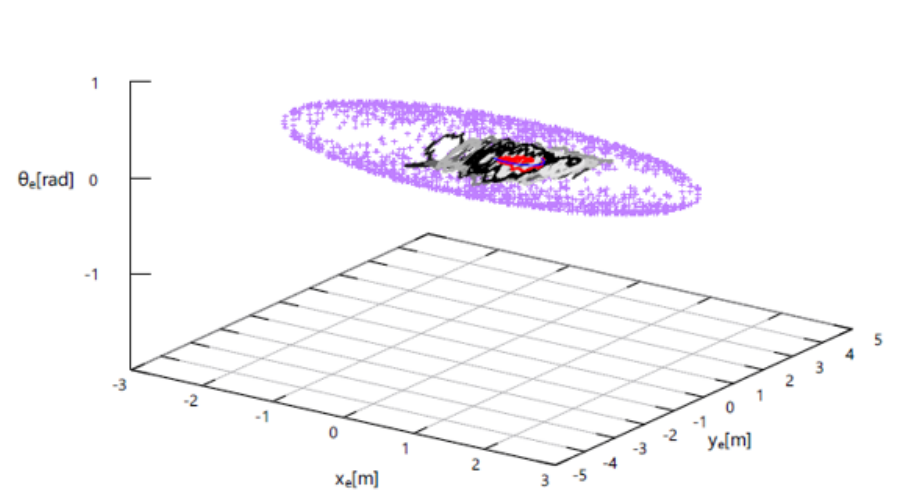}
  \caption{Trajectories with $u=u_{tra}+u_{nlc}$ and safety boundary.}
  \label{fig:withinp_nonlinear:f3a}
  \includegraphics[width=0.9\linewidth]{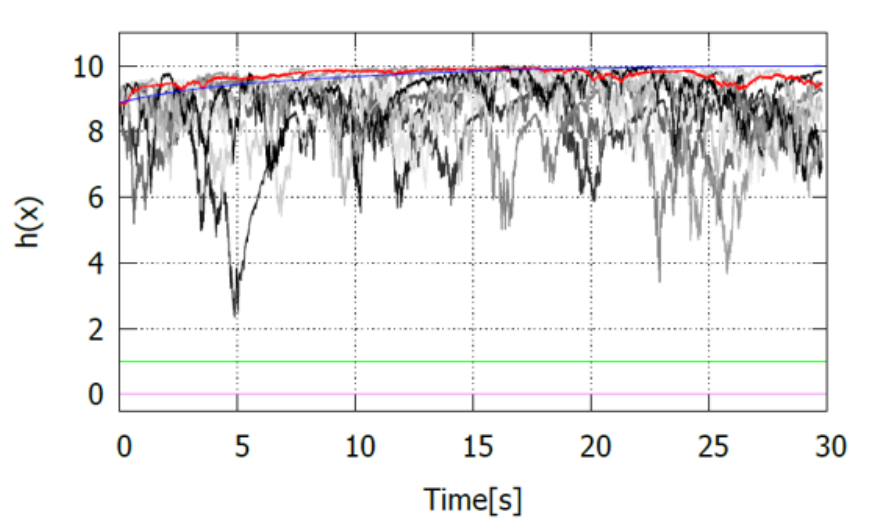}
  \caption{Time responses of $h(x)$  with $u=u_{tra}+u_{nlc}$.}
  \label{fig:withinp_nonlinear:f3b}
    \end{figure}

\begin{figure}[t]
  \centering
  \includegraphics[width=0.9\linewidth]{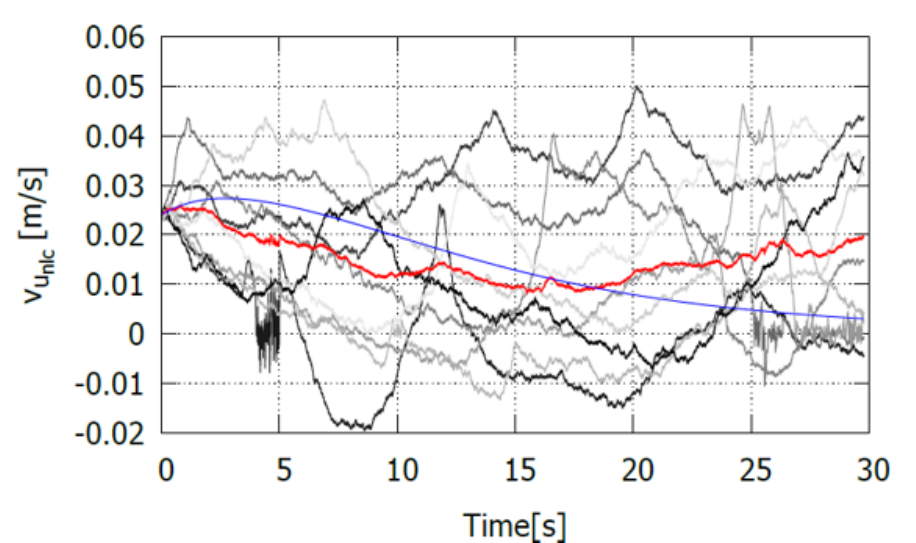}
  \caption{Time responses of $v_{nlc}$.}
  \label{fig:withinp_nonlinear:f3c}
  \includegraphics[width=0.9\linewidth]{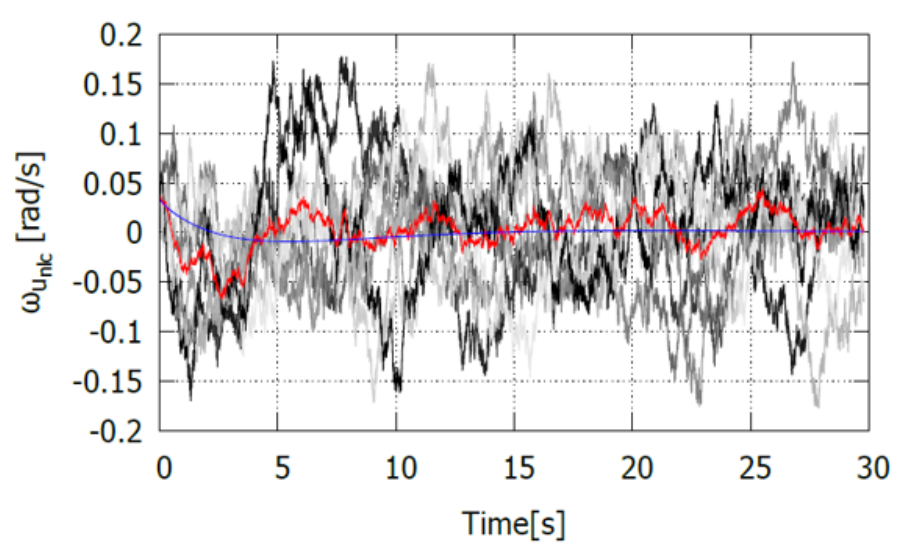}
  \caption{Time responses of $\omega_{nlc}$.}
  \label{fig:withinp_nonlinear:f3d}

\end{figure}

Figs.~\ref{fig:noinp:f1a}--\ref{fig:withinp_nonlinear:f3d} show the results of $u=u_{tra}(x)$, $u=u_{tra}(x)+u_{com}(x)$ and $u=u_{tra}(x)+u_{nlc}(x)$, respectively.  
The ten sample paths are depicted by gray lines, the average of the sample paths is shown by the red line, and the results without noise (that is, $G = 0$) are shown by the blue line. 
Moreover, the purple lines in Figs.~\ref{fig:noinp:f1a}, \ref{fig:noinp:f1b}, \ref{fig:withinp:f2a}, \ref{fig:withinp:f2b}, \ref{fig:withinp_nonlinear:f3a}, and \ref{fig:withinp_nonlinear:f3b} indicate the boundary of the safe set $\chi$, and the green line in Figs.~\ref{fig:noinp:f1b}, \ref{fig:withinp:f2b}, \ref{fig:withinp_nonlinear:f3b} indicates the boundary of the initial state set $\chi_{h>\mu}$.

When $u=u_{tra}(x)$; that is, without any safety-probability compensator, $h(x)$ takes negative values as shown in Fig.~\ref{fig:noinp:f1b}; therefore, in Fig.~\ref{fig:noinp:f1a}, the trajectories of $x$ escape from the boundary of the safe set. This result accurately reflects the theoretically calculated safety probability of $0.0043$. 

When $u=u_{tra}(x) + u_{com}(x)$; that is, with the linear safety-probability compensator, all sample paths remain in the safe set $h(x)>0$ as shown in Fig.~\ref{fig:withinp:f2b}; therefore, in Fig.~\ref{fig:withinp:f2a},the trajectories of $x$ also remain inside the safe set. This result accurately reflects the theoretically calculated safety probability of $0.997$. 

When $u=u_{tra}(x) + u_{nlc}(x)$; that is, with the nonlinear safety-probability compensator, all sample paths remain in the safe set $h(x)>0$ as shown in Fig.~\ref{fig:withinp_nonlinear:f3b}; therefore, in Fig.~\ref{fig:withinp_nonlinear:f3a}, the trajectories of $x$ also remain inside the safe set. This result accurately reflects the theoretically calculated safety probability of $0.950$.



\section{Theory}\label{sec:theory}

\subsection{Notations}
Let $\R^n$ be an $n$-dimensional Euclidean space and especially, $\R:=\R^1$. A Lie derivative of a smooth mapping $W:\R^n\to\R$ in a mapping $F=(F_1,...,F_q):\R^n\to\R^n\times\R^q$ with $F_1,...,F_q:\R^n\to\R^n$ is denoted by 
\begin{align}   L_fW(x)=\left(\frac{\partial W}{\partial x}F_1(x),...,\frac{\partial W}{\partial x}F_q(x)\right).  
\end{align}
For constants $a,b>0$, a continuous mapping $\alpha:[-b,a]\to\R$ is said to be an extended class $\mathcal{K}$ function if it is strictly increasing and satisfies $\alpha(0)=0$. A class $\mathcal{K}$ function $\alpha$ is said to be of class $\mathcal{K}_\infty$ if $\lim_{s\to\infty}\alpha(s)=\infty$. If a function $\alpha$ is $r$ times continuously differentiable, it is denoted as $C^r$. The boundary of a set $\mathcal{A}$ is denoted by $\partial\mathcal{A}$. 

Let $(\Omega,\mathcal{F},\{\mathcal{F}_t\}_{t\geq 0},\mathbb{P})$ be a filtered probability space where $\Omega$ is the sample space, $\mathcal{F}$ is the $\sigma$-algebra that is a collection of all the events, $\{\mathcal{F}_t\}_{t\geq 0}$ is a filtration of $\mathcal{F}$ and $\mathbb{P}$ is a probabilistic measure. In the filtered probability space, $\mathbb{P}[A|A_o]$ denotes the probability of some event $A$ under some condition $A_o$ and $W_t$ is a $d$-dimensional standard Wiener process. For a Markov process $X_t\in \R^n$ with an initial state $X_t=x_0$, we often use the following notation $\mathbb{P}_{x_0}[A]=\mathbb{P}[A|X_0=x_0]$. The differential form of an It\^o integral of $f:\R^n\to \R_n$ over $W_t$ is represented by $f(x)dW_t$. The trace of a square matrix $Q$ is denoted by $\tr[Q]$. 

\subsection{Stochastic System}
We consider the following stochastic system:
\begin{align}\label{system}
    dX_t=\{f(X_t)+g(X_t)(u_o(X_t)+u(t))\}dt+\sigma(X_t)dW_t, 
\end{align}
where $X_t\in\R^n$ is a state vector with a fixed initial value $x_0=x(0)\in\R^n$, $u_o:\R^n\to \R^m$ is a pre-input assumed to be a continuous state-feedback, $u\in U \subset \R^m$ is a compensator for safety-critical control, where $U$ denotes an admissible control set, and maps $f:\R^n\to \R^n$, $g:\R^n\to \R^n\times \R^m$, $\sigma:\R^n\to\R^n\times\R^d$ are assumed to be all locally Lipschitz. The local Lipschitz condition on $f$, $g$, $\sigma$ implies the existence of a stopping time $T>0$ such that $(X_t)_{t<T}$ is the maximal solution of the system. 

For a $C^2$ mapping $y:M\to\R$, with $M\subset\R^n$, 
\begin{align}
    &L_{f,g}^D(u,u_o(x),y(x)):=(L_fy)(x)+(L_gy)(x)(u+u_o(x))\\
   &L_\sigma^I(y(x)):=\frac{1}{2}\tr\left[\sigma(x)\sigma(x)^T\left[\frac{\partial}{\partial x}\left[\frac{\partial y}{\partial x}\right]^T\right](x)\right]\\
    &\mathcal{L}_{f,g,\sigma}(u,u_o,y(x)):=L_{f,g}^D(u,u_o(x),y(x))+L_\sigma^I(y(x))\\
    &H_{\sigma}(h(x)):=\frac{1}{2}L_\sigma h(x)(L_\sigma h(x))^T. 
\end{align}

\subsection{Safety-probability Analysis and Design}

\subsubsection{Safety Probability}
In this subsection, we improve the stochastic safety analysis method proposed in \cite{nishimura:24}, which is based on stochastic zeroing control barrier functions (stochastic ZCBFs). 

Let us define a safe set $\chi \subset \mathbb{R}^n$ being open, and there exists a mapping $h: \mathbb{R}^n \to \mathbb{R}$ satisfying all the following conditions:
\begin{enumerate}
    \item [$\mathbf{(Z1)}$] $h(x)$ is $C^2$. 
    \item [$\mathbf{(Z2)}$]$h(x)$ is proper on $\chi$; that is, for any $L > 0$, for any superlevel set $\{x \in \R^n | h(x) \ge L\}$ is compact. 
    \item [$\mathbf{(Z3)}$]The closure of $\chi$ is the 0-superlevel set of $h(x)$; that is, 
    \begin{align}
        \chi&=\{ x\in \R^n|h(x)>0\},\\
        \partial \chi &=\{ x\in\R^n|h(x)=0\}
    \end{align}
     are both satisfied. 
\end{enumerate}

   We set some sets and stopping times used in this subsection. For $\mu>0$, let
\begin{align}
    &\chi_\mu:=\{x\in\R^n|h(x)\in(0,\mu]\}\subset\chi\\
   &\chi_{h>\mu}:=\chi\backslash\chi_\mu=\{x\in\R^n|h(x)>\mu\}\\
   &\R^n_{h\leq \mu}:=\tilde{\chi}\cup{\chi}_\mu=\{x\in \R^n|h(x)\leq \mu\}
\end{align}
   be defined. For a solution to system \eqref{system} with $x_0 \in \chi$, the first exit time from $\chi$ is denoted by $\tau_0$.  

Let $p\in [0,1]$. System \eqref{system} said to be \emph{safe in $(\chi_{h > \mu},\chi,p)$} if, for any $x_0\in \chi_{h > \mu}$, 
    \begin{align}
    \mathbb{P}_{x_0} \left[ \inf_{t \ge 0} h(X_{t \wedge \tau_0}) > 0 \right] \ge p
\end{align}
    is satisfied. 

To analyze the safety of a stochastic system, 
we  define the following:
\begin{definition}\label{definition}
(Stochastic ZCBF \cite{nishimura})
Let \eqref{system} be considered with $\chi$ and $h(x)$ satisfying (Z1), (Z2) and (Z3). If there exist continuous mapping $\phi:\R^n\to\R^m$ and $b>0$ such that, for all $x\in\R^n_{h\leq \mu}$, 
\begin{align}
    \label{eq:con-prob}  \mathcal{L}_{f,g,\sigma}(u_o(x),\phi(x),h(x))\geq bH_\sigma(h(x))
\end{align}
is satisfied with some $b>0$, then $h(x)$ is said to be \emph{a stochastic zeroing control barrier function (ZCBF)}. 
\end{definition}
\begin{theorem}\label{THM:SAFETY}
Let the system \eqref{system} be considered. If there exists a stochastic ZCBF $h(x)$, then the system becomes safe in $(\chi_{h>\mu},\chi,1-e^{-b\mu})$ by designing $u=\phi(x)$ that satisfies all the conditions in Definition~\ref{definition}. \eot
\end{theorem}
The proof for Theorem~\ref{THM:SAFETY} is written in Appendix~\ref{app:safety} below.

\begin{remark}
    The results of this subsection are different from those in \cite{nishimura:24} in two points. The first is that we define safe in $(\chi_{h>\mu}, \chi,p)$, which is the notion for a global time; in contrast, the previous work considers {\it transiently safe}, which is the notion for a specific Markov time. The other is that we set the initial state set (the first element of the triple for safety) as $\chi_{h>\mu}$, while the previous work considers the initial state set as $\chi_\mu$. The change in the initial state enables us to consider safety in global time. \eor
\end{remark}

\subsubsection{Safety Probability Analysis and Linear Control Design For Stochastic Linear Systems}
In this subsection, we improve the design compensators of the safety probability for a linear system with additive noises by improving the results of \cite{nishimura} using the safety proposed in the previous subsection.

Consider a linear system with additive noise
\begin{align}\label{dx_1}
    dX_t = (AX_t + B u_{o}(X_t) )dt+ GdW_t, 
\end{align}
which is \eqref{system} with $f(x)=Ax$, $g(x)=B$, $\sigma(x)=G$ and $u=0$, where $A\in\mathbb{R}^{n\times n} $, $B\in \R^{n\times m}$ and $G\in\mathbb{R}^{n\times d} $. 
\begin{theorem}\label{THM:LINEAR}
   Letting $u_o(x) = -BK x$ with $K \in \R^{m \times n}$ and $\bar{A}=A-BK$, assume that there exist positive definite and symmetric matrices $P, Q\in\mathbb{R}^{n\times n}$ satisfying a Lyapunov equation
\begin{align}
P \bar{A} + \bar{A}^T P = -Q. 
\end{align}
Let us also consider a candidate for a stochastic ZBF
\begin{align}
\label{eq:zcbf-lin} h(x) = -x^T P x + M, \quad M > 0, 
\end{align}
a safe set $\chi$ and the related sets $ \chi_\mu$, $\chi_{h>\mu}$, and  $\mathbb{R}^n_{h\leq\mu}$ with $ \mu \in (0, M) $. If
\begin{align}\label{L}
L := \mathrm{eigmin}[Q] - \mathrm{eigmin}[P]\frac{  \mathrm{tr}[G^T P G]}{M - \mu} > 0
\end{align}
is satisfied, then the system \eqref{dx_1} is safe in $(\chi_{h>\mu}, \chi, 1 - e^{-b\mu})$, where 
\begin{align}\label{Lb}
b \leq \frac{L}{2\mathrm {eigmax} [PGG^TP]}. 
\end{align}
\eot
\end{theorem}
The proof for Theorem~\ref{THM:LINEAR} is written in Appendix~\ref{app:linear} below.

Next, we consider adding the compensator $u$ for \eqref{dx_1}; that is,
\begin{align}\label{dx_2}
    dX_t = (AX_t + B( u_o(X_t) + u(t)) ) dt+G dW_t.
\end{align}

\begin{corollary}\label{COR:SAFE-LIN-CLO2}
Assume that the conditions in Theorem~\ref{THM:LINEAR} are all satisfied. If there exist a positive definite and symmetric matrix $R \in \R^m \times \R^m$ and $b^{+}>0$ such that
\begin{align}
    B R^{-1} B^T = b^{+}G G^T,  \label{eq:con-pseudo-opt}
\end{align}
then, the system \eqref{dx_2} with $u=\phi_{po}(x)$, where
\begin{align}
    \phi_{po}(x) := -R^{-1}B^TPx,  \label{eq:ctrl-pseudo-opt}
\end{align}
is safe in $(\chi_{h>\mu},\chi,1-e^{-(b+b^{+})\mu})$. \eot
\end{corollary}
The proof for Corollary~\ref{COR:SAFE-LIN-CLO2} is shown in Appendix~\ref{app:COR:SAFE-LIN-CLO2}.

\subsubsection{Safety Probability Control Design For Stochastic Nonlinear Systems}
In this subsection, we provide the design procedure for a compensator of the safety probability for a nonlinear system by applying our result of Theorem~\ref{THM:SAFETY} to the result of \cite{nishimura:24}. 

\begin{corollary}\label{COR:CTRL-SZCBF}
    Let the system \eqref{system} be considered with the safe set $\chi$ and a candidate of a stochastic ZCBF $h(x)$ satisfying all the conditions of (Z1)--(Z3). Let
    \begin{align}
        &I_s (u_o(x),h(x)) := \mathcal{L}_{f,g,\sigma}(0,u_o(x),h(x)) \\
        &J_s(h(x)) := b H_\sigma(h(x))
    \end{align}
and $\phi_s: \R^n \to \R^m$ be designed as 
    \begin{align}
        \phi_s(x) = 
            -\frac{I_s(u_o(x),h(x))-J_s(h(x))}{L_gh(x) (L_gh(x))^T} (L_gh(x))^T
    \end{align}
    for $I_s < J_s \cap L_gh \neq 0$, and $\phi_s(x)=0$ for $I_s \ge J_s \cup L_gh = 0$. Moreover, we consider
\begin{align}
    \Phi_s := \left\{ \begin{array}{ll}
        \phi_s, & x \in \R^n_{h\le \mu},\\
        \phi'_s, & x \in \chi_{h>\mu},
        \end{array}\right.
\end{align}
where $\phi'_s: \chi_{h>\mu} \to \R^m$ is continuous and satisfies $\phi'_s(x) = \phi_s(x)$ for all $x \in \partial \chi_{h > \mu}$.

If, for all $x \in \chi_\mu$ with some $\mu > 0$ satisfying $L_gh=0$, 
    \begin{align}
    \label{eq:ctrl-con-prob}    L_fh(x) + L^I_\sigma(h(X)) > b H_\sigma(h(x))
    \end{align}
holds, then $\Phi_s$ is continuous all in $\R^n$ and the system \eqref{system} with $u=\Phi_s(x)$ is safe in $(\chi_{h > \mu},\chi,1-e^{-b\mu})$. \eot 
\end{corollary}
While the proof is quite similar to Corollary~2 in \cite{nishimura:24}, we describe it Appendix~\ref{app:COR:CTRL-SZCBF} for the sake of self-containment in this paper.

\section{Conclusion}\label{sec:conclusion}

In this paper, we modified stochastic safety-critical control theory using the stochastic zeroing control barrier function (ZCBF) in \cite{nishimura} and applied it to safety-critical compensation for a trajectory tracking problem of marine vessels subject to irregular disturbances. The error dynamics of the vessel motion is stated in the state-space model and the tracking control is achieved by linear quadratic control for the linearized model. The safety probability was defined as the probability that the error trajectory remains the designed region against irregular disturbances, which is assumed to be Gaussian white noise. Then, the linear and nonlinear safety probability compensators were proposed based on the modified stochastic safety-critical control theory, and the effects of the compensators were confirmed by numerical simulation.

\bibliographystyle{plain}
\bibliography{ifacconf}

@book{fossen,
	author={T.~I.~Fossen},
	title={Handbook of Marine Craft Hydrodynamics and Motion Control},
    edition={Second},
	publisher={John Wiley \& Sons Ltd.},
    address = {Chichester, UK},
	year={2021},
}

@article{fujii,
	author={Y. Fujii and H. Nakamura and Y. Sato},
	title={Stability gain design method based on $\mathcal{L}_2$ norms for differentially flat systems},
	journal={Transactions of the Society of Instrument and Control Engineers},
	year={2020},
	volume={56},
    number={5},
    pages={259--268},
}

@article{saback,
	author={R.~M.~Saback and A.~G.~S.~Conceicao and T.~L.~M.~Santos and J.~Albiez and M.~Reis},
	title={Nonlinear model predictive control applied to an autonomous underwater vehicle},
	journal={IEEE J. Oceanic Engineering},
	year={2020},
	volume={45},
    number={3},
    pages={799--812},
}

@article{ames2019,
	author={A.~D.~Ames and S.~Coogan and M.~Egerstedt and G.~Notomista and K.~Sreenath and P.~Tabuada},
	title={Control barrier functions: theory and applications},
	journal={Proc. 18th Euro. Control Conf.},
	year={2019},
    pages={3420--3431},
}

@article{otsuki,
	author={S.~Otsuki and N.~Hatta and M.~Hanif and T.~Hatanaka and K.~Nakashima},
	title={Hierarchical vessel autonomous operation in a port with safety certificates: combined MPC and CBF approach},
	journal={Proc. IFAC World Congress 2023},
	year={2023},
    pages={3481--3488},
}

@article{prajana,
	author={S.~Prajna and A.~Jadbabaie and G.~J.~Pappas},
	title={A framework for worst-case and stochastic safety verification using barrier certificates},
	journal={IEEE Trans. Autom. Control},
	year={2007},
	volume={52},
    number={8},
    pages={1415--1428},
}

@article{xue,
	author={B.~Xue and N.~Zhan and M.~ Franzle},
	title={Reach-avoid analysis for stochastic differential equations},
	journal={IEEE Trans. Autom. Control},
	year={2024},
	volume={69},
    number={3},
    pages={1882--1889},
}

@article{nejati,
	author={A.~Nejati and S.~Soudjani and M.~Zamani},
	title={Compositional construction of control barrier functions for continuous-time stochastic hybrid systems},
	journal={Automatica},
	year={2022},
	volume={145},
    pages={110513},
}

@article{nishimura,
	author={Y.~Nishimura and K. Hoshino},
	title={Safety-Probability Analysis and Control for Stochastic Systems Based on Lyapunov Candidate Functions},
	journal={Proc. 62nd IEEE Conf. Decis. Contr.},
	year={2023},
    pages={4818--4823},
}

@article{nishimura:24,
    author={Y.~Nishimura and K.~Hoshino},
    title={Control Barrier Functions for Stochastic Systems and Safety-Critical Control Designs},
    journal={IEEE Trans. Autom. Control},
	year={2024},
	volume={69},
    number={11},
    pages={8088--8095},
}

@article{esfahani2021,
    author={H.~N.~Esfahani and R.~Szlapczynski},
    title={Robust-adaptive dynamic programming-based time-delay control of autonomous ships under stochastic disturbances using an actor-critic learning algorithm},
    journal={J Marine Sci Technol},
	year={2021},
	volume={26},
    number={4},
    pages={1262--1279},
}

@article{maki2023,
    author={A.~Maki and K.~Hoshino and L.~Dostal and Y.~Maruyama and F.~Hane and Y.~Yoshimura},
    title={Stochastic stabilization and destabilization of ship maneuvering motion by multiplicative noise},
    journal={J Marine Sci Technol},
	year={2023},
	volume={28},
    number={8},
    pages={704--718},
}

@article{maki2024,
    author={A.~Maki and Y.~Maruyama and Y.~Liu and L.~Dostal},
    title={Comparison of stochastic stability boundaries for parametrically forced systems with application to ship rolling motion},
    journal={J Marine Sci Technol},
	year={2024},
	volume={29},
    number={5},
    pages={446--456},
}

\appendix






\section{Proof of Theorem~\ref{THM:SAFETY}}\label{app:safety}    

\subsection{Existence of Solution}
First, we prove that the existence of a stochastic ZCBF $h(x)$ ensures that the system \eqref{system} with $u=\phi(x)$ has a solution in global time. To do this, we consider the following definition and theorem.

\begin{definition}[FCiP, \cite{nishimura}]\label{def:fcip}
Let system \eqref{system} be considered with $u=\phi(x)$, where $\phi: \R^n \to \R^m$ is a continuous mapping. If a $C^2$ mapping $Y: \R^n \to [0,\infty)$ is proper; that is, for any $L \in [0,\infty)$, any sublevel set $\{x \in \R^n | Y(x) \le L\}$ is compact, and a continuous mapping $\psi: [0,\infty) \times (0,1) \to [0,\infty)$ both exist for every $x_0 \in \R^n$ such that
\begin{align}\label{eq:fcip}
\pri{x_0}{\forall t \in [0,l],\ Y({X_t}) \le \psi(l,\epsilon)} \ge 1- \epsilon
\end{align}
holds for all $l \in [0,\infty)$ and all $\epsilon \in (0,1]$, then the system is said to be forward complete in probability (FCiP). \eod
\end{definition}

\begin{theorem}{(\cite{nishimura})}\label{thm:fcip}
Let us consider system \eqref{system}, a continuous mapping $\phi: \R^n \to \R^m$ and an initial condition $x_0 \in \R^n$. If there exists a proper and $C^2$ mapping $Y: \R^n \to [0,\infty)$ such that
\begin{align}
\mathcal{L}_{f,g,\sigma}(\phi(x),u_o(x),Y(x)) \le c_1 Y(x) + c_2
\end{align}
is satisfied for all $x \in \R^n$ and for some $c_1 \in [0,\infty)$ and $c_2 \in [0,\infty)$, then the system with $u=\phi(x)$ is FCiP. \eot
\end{theorem}

Let
\begin{align}\label{eq:func-prob}
h_b(x) := e^{bh(x)}.
\end{align}
Because 
\begin{align}
L^D_{f,g}(\phi(x),u_o(x),h_b(x)) = b h_b(x) L^D_{f,g}(\phi,u_o(x),h(x)),
\end{align}
is satisfied, \eqref{eq:con-prob} changes as follows:
\begin{align}\label{eq:con-prob3} 
L^D_{f,g}(\phi(x),u_o(x),h_b(x)) \ge b h_b(x) \left\{ b H_{\sigma}(h(x)) - L^I_{\sigma}(h(x))\right\}.
\end{align}
Moreover, letting
\begin{align}
B_b(x):=(h_b(x))^{-1}=e^{-bh(x)},
\end{align}
we obtain
\begin{align}\label{eq:htob}
L^I_{\sigma}(B_b(x)) = b B_b(x) \left\{ b H_{\sigma}(h(x)) - L^I_{\sigma}(h(x)) \right\},
\end{align}
which transforms \eqref{eq:con-prob3} into
\begin{align}\label{eq:con-prob4} 
L^D_{f,g}(\phi,u_o(x),h_b(x)) \ge (h_b(x))^2 L^I_{\sigma}(B_b(x)).
\end{align}
Therefore, using the relationship
\begin{align}\label{eq:rel-bh}
L^D_{f,g}(u,u_o(x),h_b(x)) = -(h_b(x))^{2} L^D_{f,g}(u,u_o(x),B_b(x)),
\end{align}
we obtain
\begin{align}
-L^D_{f,g}(\phi,u_o(x),B_b(x)) \ge L^I_{\sigma}(B_b(x));
\end{align}
that is, 
\begin{align}\label{eq:exp-szcbf}
\mathcal{L}_{f,g,\sigma}(\phi,u_o(x),B_b(x)) \le 0,\ x \in \R^n_{h \le \mu}.
\end{align}

Here, we consider the rest space $\chi_{h > \mu}$, where the assumption (Z2) implies that the space is bounded and $h$ is bounded from above in the space. In addition, $B_b$ is decreasing, $u_o$ is continuous, and $f$, $g$, and $\sigma$ are all locally Lipschitz. Therefore, $\mathcal{L}_{f,g,\sigma}(\phi,u_o(x),B_b(x))$ is bounded from above; that is, for sufficiently large values $c_1>0$ and $c_2>0$, we obtain
\begin{align}\label{eq:exp-szcbf-soto}
\mathcal{L}_{f,g,\sigma}(\phi,u_o(x),B_b(x)) \le c_1 B_b(x) + c_2,\ x \in \chi_{h > \mu}.
\end{align}
Considering \eqref{eq:exp-szcbf} and \eqref{eq:exp-szcbf-soto}, all the conditions of Theorem~\ref{thm:fcip} are satisfied with $Y=B_b$; that is, the system \eqref{system} with $u=\phi(x)$ is FCiP.

\subsection{Safety Probability}
Next, considering
\begin{align}
	\{ X_t \in \chi, \forall t \ge 0 \}		&=\{ \tau_0 = \infty \} \\
		&=\{ \inf_{t \ge 0} h(X_t) > 0 \} \\
		&=\{ \inf_{t \ge 0} h(X_{t \wedge \tau_0}) > 0\} \\
		&=\{ \sup_{t \ge 0} B_b(X_{t \wedge \tau_0}) < 1 \}
\end{align}
and Markov inequality, for $x \in \chi_{h>\mu}$, 
\begin{align}
	&\pri{x_0}{ \{ \sup_{t \ge 0} B_b(X_{t \wedge \tau_0}) \ge m\} \cap \{ \tau_\mu < \infty\} } \nonumber \\
    &\quad \le \frac{ \exi{x_0}{ \sup_{t \ge 0}B_b(X_{t \wedge \tau_0}) I_{\{ \tau_\mu < \infty\} }}}{m}
\end{align}
for any $m>0$. Choosing $m=1$, we obtain
\begin{align}
	&\pri{x_0}{ \{\inf_{t \ge 0} h(X_{t \wedge \tau_0}) \le 0 \} \cap \{ \tau_\mu < \infty\} } \nonumber \\
    &\quad = \pri{x_0}{ \{\sup_{t \ge 0}B_b(X_{t \wedge \tau_0}) \ge 1\} \cap \{ \tau_\mu < \infty\} } \\
	&\quad \le \exi{x_0}{\sup_{t \ge 0}B_b(X_{t \wedge \tau_0}) I_{\{ \tau_\mu < \infty\} }}. \label{eq:jouken2}
\end{align}

Here, considering the strong Markov property, we obtain
\begin{align}
	&\exi{x_0}{\sup_{t \ge 0} B_b(X_{t \wedge \tau_0}) I_{\{ \tau_\mu < \infty \}}} \nonumber \\
    &\quad= \exi{x_0}{ \exi{x_{\tau_\mu}}{\sup_{t \ge \tau_\mu} B_b(X_{(t-\tau_\mu) \wedge \tau_0})} I_{\{ \tau_\mu < \infty \} }} \\
	&\quad \le \exi{x_0}{B_b(X_{\tau_\mu}) I_{\{ \tau_\mu < \infty \} }} \\
	&\quad = e^{-b\mu} \pri{x_0}{ \{ \tau_\mu < \infty\} },
\end{align}
where we use Dynkin's formula with the given condition \eqref{eq:exp-szcbf}:
\begin{align}
	&\exi{x_{\tau_\mu}}{B_b(X_{(t-\tau_\mu)\wedge \tau_0})} - B_b(X_{\tau_\mu}) \\
	&= \exi{x_{\tau_\mu}}{\int_{\tau_\mu}^{(t-t\mu)\wedge \tau_0} \mathcal{L}_{f,g,\sigma}(\phi(X_\tau),u_o(X_\tau),B_b(X_\tau) d\tau} \nonumber \\
    &\le 0.
\end{align}
Substituting the above result to \eqref{eq:jouken2}, we obtain
\begin{align}
	&\pri{x_0}{ \{\inf_{t \ge 0} h(X_{t \wedge \tau_0}) \le 0 \} \cap \{ \tau_\mu < \infty\} } \nonumber \\
    &\quad \le e^{-b\mu} \pri{x_0}{ \{ \tau_\mu < \infty\} };
\end{align}
thus, 
\begin{align}
	&\pri{x_0}{\{ X_t \in \chi, \forall t \ge 0 \} \cap \{\tau_\mu < \infty \}} \nonumber \\
    &\quad = \pri{x_0}{ \{\inf_{t \ge 0}h(X_{t \wedge \tau_0}) > 0 \} \cap \{ \tau_\mu < \infty\} } \\
	&\quad \ge (1 - e^{-b\mu}) \pri{x_0}{ \{ \tau_\mu < \infty\} } \label{eq:jouken3}
\end{align}

On the other hand, for $x \in \chi_{h>\mu}$, we obtain
\begin{align}
	&\pri{x_0}{ \{ X_t \in \chi, \forall t \ge 0 \} \cap \{ \tau_\mu = \infty \} } \nonumber \\
    &\quad = \pri{x_0}{\{ \tau_\mu = \infty \}} \\
	&\quad \ge (1 - e^{-b \mu}) \pri{x_0}{\{ \tau_\mu = \infty \}}	\label{eq:jouken1}
\end{align}

Combining \eqref{eq:jouken1} and \eqref{eq:jouken3}, we obtain
\begin{align}
	&\pri{x_0}{ \{ X_t \in \chi, \forall t \ge 0 \}} \nonumber \\
    &\quad = \pri{x_0}{ \{ X_t \in \chi, \forall t \ge 0 \} \cap \{ \tau_\mu = \infty\} } \nonumber \\
	&\quad + \pri{x_0}{ \{ X_t \in \chi, \forall t \ge 0 \} \cap \{ \tau_\mu < \infty\} } \\
	&\quad \ge (1 - e^{-b \mu}) \left( \pri{x_0}{\{ \tau_\mu = \infty \}} + \pri{x_0}{\{ \tau_\mu < \infty \}} \right) \\
	&\quad =1 - e^{-b \mu}.
\end{align}
This completes the proof.

\section{Proof of Theorem~\ref{THM:LINEAR}} \label{app:linear}             

This theorem is proven by showing that $h(x)$ is a stochastic ZCBF. In the proof, we often use the relationship
\begin{align}
    \mathrm{eigmin}[Y] x^T x \le x^T Y x \le \mathrm{eigmax}[Y] x^T x \label{eq:eigrelation}
\end{align}
for a symmetric matrix $Y \in \R^n \times \R^n$. 

First, because the assumption of \eqref{eq:con-prob} has to hold for $x \in \R^n_{h \le \mu}$; that is,
\begin{align}
    x^T P x \ge M - \mu.
\end{align} 
Therefore, it is sufficient for
\begin{align}
    x^T x \ge \frac{M-\mu}{\mathrm{eigmin}[P]} \label{eq:pr-hmu}
\end{align}
to satisfy assumption \eqref{eq:con-prob}. On the other hand, the given assumptions \eqref{L} and \eqref{Lb} yield
\begin{align}
    \mathrm{eigmin}[Q] - 2b \cdot \mathrm{eigmax}[PGG^TP] \ge  \mathrm{eigmax}[P] \frac{\mathrm{tr}[G^T P G]}{M-\mu}.
\end{align}
Thus, we obtain the following condition of
\begin{align}\label{eq:givencon-result}
    \frac{M-\mu}{\mathrm{eigmin}[P]} \ge \frac{\mathrm{tr}[G^T P G]}{\mathrm{eigmin}[Q] - 2b \cdot \mathrm{eigmax}[PGG^TP]}.
\end{align}
This results in
\begin{align}
    x^T x \ge \frac{\mathrm{tr}[G^T P G]}{\mathrm{eigmin}[Q] - 2b \cdot \mathrm{eigmax}[PGG^TP]}.
\end{align}
Applying \eqref{eq:eigrelation} to the above inequality, we obtain
\begin{align}
    x^T Q x -\mathrm{tr}[G^T P G] \ge 2b x^T PGG^T P x,
\end{align}
which is the same as 
\begin{align}
    \mathcal{L}_{Ax,B,G}(0,-BKx,h(x)) \ge b H_G(h(x)). \label{eq:safe-pr-L}
\end{align}
This is a sufficient condition that $h(x)$ is a stochastic ZCBF with $\phi=0$. Consequently, by Theorem~\ref{THM:SAFETY}, the system is safe in $(\chi_{h>\mu},\chi,1-e^{-b\mu})$. 

\section{Proof of Corollary~\ref{COR:SAFE-LIN-CLO2}}\label{app:COR:SAFE-LIN-CLO2}

Considering $u=\phi_{po}(x)$ with \eqref{eq:ctrl-pseudo-opt} and $h(x)$ with \eqref{eq:zcbf-lin}, we obtain
\begin{align}
    \mathcal{L}_{Ax,B,G}(\phi_{po}(x),-BKx,h(x)) &= x^T Q x -\mathrm{tr}\left[ G^T P G \right] \nonumber \\
        &\quad + 2 x^T P B R^{-1} B^T P x;
\end{align}
thus, applying \eqref{eq:safe-pr-L}, we obtain
\begin{align}
    \mathcal{L}_{Ax,B,G}&(\phi_{po}(x),-BKx,h(x)) \nonumber \\
        &\ge 2b x^T PGG^T P x + 2 x^T P B R^{-1} B^T P x.
\end{align}
Moreover, we also consider the additional assumption \eqref{eq:con-pseudo-opt}, the above inequality results in
\begin{align}
    \mathcal{L}_{Ax,B,G}(\phi_{po}(x),-BKx,h(x)) &\ge 2 (b + b^{+}) x^T PGG^T P x,
\end{align}
which implies \eqref{eq:con-prob}, provided that $b$ is replaced by $b+b^{+}$. This completes the proof.

\section{Proof of Corollary~\ref{COR:CTRL-SZCBF}}\label{app:COR:CTRL-SZCBF}
First, consider the case $\lie{g}{h} \neq 0$ in $\chi_\mu$. If $I_s < J_s$, we obtain
\begin{align}
\mathcal{L}_{f,g,\sigma}&(\phi_s(x),u_o(x),h(x)) = b H_\sigma(h(x))
\end{align}
and if $I_s \ge J_s$, we obtain
\begin{align}
\mathcal{L}_{f,g,\sigma}(\phi_s(x),u_o(x),h(x)) &= I_s(u_o(x),h(x)) \nonumber \\
&\ge J_s(h(x)) \nonumber = b H_\sigma(h(x)).
\end{align}
Therefore, regardless of $I_s < J_s$ or $I_s \ge J_s$, the inequality \eqref{eq:con-prob} is satisfied. Moreover, because $\lie{g}{h(x)}$, $I_s(u_o(x),h(x))$ and $J_s(h(x))$ are all continuous in $\lie{g}{h(x) \neq 0}$ and $\phi_s(x) \to 0$ as $I_s \to J_s$ uniformly when $\lie{g}{h(x)} \neq 0$, $\phi_s(x)$ is continuous in $\lie{g}{h(x)} \neq 0$. 
Then, we consider the other case, i.e., $\lie{g}{h} = 0$ in $\chi_\mu$. 
The additional condition \eqref{eq:ctrl-con-prob} implies that there exists a sufficiently small constant $\epsilon>0$ such that
\begin{align}
\lie{f}{h}(x) + L^I_\sigma(h(x)) - \epsilon \ge b H_\sigma(h(x))
\end{align}
is satisfied. Combining the inequality and the assumption of $u_o$ to be continuous, for a subset $G_{o\mu} \subset \chi_\mu$, which is a neighborhood of $x_g \in \{ x \in \chi_\mu | \lie{g}{h}(x)=0 \}$ \begin{align}
||\lie{g}{h}(x) u_o(x) || \le \epsilon
\end{align}
is satisfied. Thus, for $x \in G_{o\mu}$, we obtain 
\begin{align}
\lie{f}{h}(x) + L^I_\sigma(h(x)) + \lie{g}{h}(x) u_o(x) \ge b H_\sigma(h(x)),
\end{align}
which implies that $I_s \ge J_s$; namely, $\phi_s=0$ in $G_o$. Therefore, $\phi_s$ is continuous around $\lie{g}{h}(x)=0$ in $\chi_\mu$. 

Consequently, $\phi_s$ is always continuous in $\chi$ and satisfies all the assumptions and conditions of Theorem~\ref{THM:SAFETY}. Moreover, because $u=\phi'_s(x)$ is continuous in $\chi_{h>\mu}$ and $\phi'_s(x)=\phi_s(x)$ for all $x \in \partial \chi_{h>\mu}$, $u$ is continuous for all $\chi$. This completes the proof.


\end{document}